%% file: MultiFuzz.tex
\documentclass[conference]{IEEEtran}

\usepackage{tikz}
\usepackage{amsmath}
\usepackage{cite}
\usepackage{amsmath,amssymb,amsfonts}
\usepackage{graphicx}
\usepackage{textcomp}
\usepackage{xcolor}
\usepackage{algorithm}
\usepackage{algpseudocode}
\usepackage{graphicx}
\usepackage{url}
\usepackage{filecontents}
\usepackage{hyperref}

\newcommand{\parabf}[1]{\paragraph*{\textbf{#1}}}

\def\BibTeX{{\rm B\kern-.05em{\sc i\kern-.025em b}\kern-.08em
    T\kern-.1667em\lower.7ex\hbox{E}\kern-.125emX}}

\begin{document}

\date{}


\title{FuzzSense: Towards A Modular Fuzzing Framework for Autonomous Driving Software}

\makeatletter
\newcommand{\linebreakand}{%
  \end{@IEEEauthorhalign}
  \hfill\mbox{}\par
  \mbox{}\hfill\begin{@IEEEauthorhalign}
}
\makeatother

\author{\IEEEauthorblockN{Andrew Roberts\IEEEauthorrefmark{1}}
\IEEEauthorblockA{
\IEEEauthorrefmark{1} FinEst Centre for Smart Cities,\\ Tallinn University of Technology
}
\and
\IEEEauthorblockN{Lorenz Teply, Mohammad Hamad, Sebastian Steinhorst\IEEEauthorrefmark{2}}
\IEEEauthorblockA{
\IEEEauthorrefmark{2}Department of Computer Engineering,\\ Technical University of Munich 
}
\and
\IEEEauthorblockN{Mert D. Pesé\IEEEauthorrefmark{3}}
\IEEEauthorblockA{
\IEEEauthorrefmark{3}School of Computing,\\ Clemson University 
}
\and
\IEEEauthorblockN{Olaf Maennel\IEEEauthorrefmark{4}}
\IEEEauthorblockA{
\IEEEauthorrefmark{4}School of Computer and Mathematical Sciences,\\ The University of Adelaide 
}
}

\maketitle

\thispagestyle{plain}
\pagestyle{plain}

\begin{abstract}
Fuzz testing to find semantic control vulnerabilities is an essential activity to evaluate the robustness of autonomous driving (AD) software. Whilst there is a preponderance of disparate fuzzing tools that target different parts of the test environment, such as the scenario, sensors, and vehicle dynamics, there is a lack of fuzzing strategies that ensemble these fuzzers to enable concurrent fuzzing, utilizing diverse techniques and targets. This research proposes \textit{FuzzSense}, a modular, black-box, mutation-based fuzzing framework that is architected to ensemble diverse AD fuzzing tools. To validate the utility of FuzzSense, a LiDAR sensor fuzzer was developed as a plug-in, and the fuzzer was implemented in the new AD simulation platform AWSIM and Autoware.Universe AD software platform. The results demonstrated that FuzzSense was able to find vulnerabilities in the new Autoware.Universe software. We contribute to FuzzSense open-source with the aim of initiating a conversation in the community on the design of AD-specific fuzzers and the establishment of a community fuzzing framework to better target the diverse technology base of autonomous vehicles.
\end{abstract}

\section{Introduction}
 Fuzz testing of autonomous driving (AD) software aims to use unsanitized and invalid input to trigger exceptional or abnormal behavior of the driving logic. AD fuzzers are designed in a disparate manner, seeding input from either the sensor data, vehicle dynamics data, scenario and simulator configuration. EnFuzz~\cite{chen2019enfuzz} demonstrated that a collective framework could ensemble diverse fuzzers exhibiting different fuzzing techniques to obtain deeper penetration of one specific type of target, in this instance, application software. As the architecture of AD software relies on a mixture of different sensor technologies and data sources, the innovation required of ensemble fuzzing for AD software is that the framework must be extensible to allow different fuzzers for different targets and target groups. Our idea with this research is to explore such a concept as an ensemble fuzzing framework for AD software and present our ideas on how this could be architected. To this end, we present \textit{FuzzSense} (\figurename~\ref{fig:HL-Architecture}), a conceptual architecture based on a modular, black-box, mutation-based fuzzing framework. 
 
The architecture of FuzzSense is envisioned to integrate within the AD software simulation environment (CARLA, AWSIM, Apollo), allowing diverse fuzzing tools as plug-ins to generate test cases, collect output data in a seed corpus, and mutate new inputs. Our motivation in presenting this work is to provoke discussion within the community on how AD systems are fuzzed, establish community efforts for fuzzing and to gather initial feedback on FuzzSense and understand potential improvements on the foundations of the design of the framework. This work is not a benchmarking study or a statistical evaluation of fuzzing performance, as the motivation is purely to understand how the design of an overarching fuzzing framework for AD software may be achieved. Therefore, to clearly state the contributions of this work, we have focused on the development of the initial concept of the AD ensemble fuzzing framework, developed source code, and conducted an initial test case. 


\begin{figure}[h!]
    \centering
\includegraphics[width=0.35\textwidth]{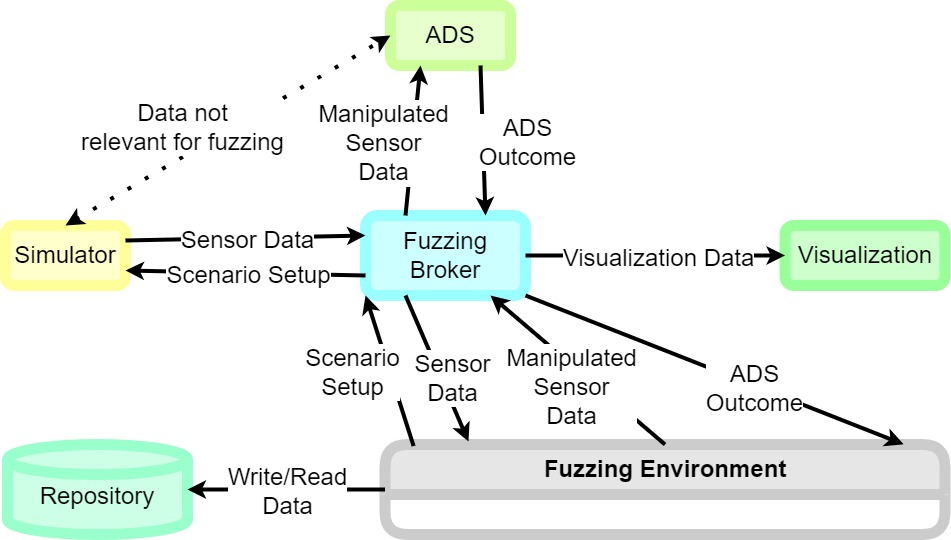}
    \caption{High-level Architecture all Components}
    \label{fig:HL-Architecture}
\end{figure}

At a more detailed level, our contributions are listed as the following: 

\begin{itemize}
    \item We present FuzzSense, an ensemble fuzzing framework for AD software.
    \item We develop a sensor fuzzing plugin for FuzzSense for the LiDAR sensor from reverse engineering LiDAR sensor configurations and applied it within AWSIM, which uses more advanced LiDAR representation technology (Rings) than popularly used CARLA.
    \item We demonstrate an initial test case of a FuzzSense plug-in to find vulnerabilities in state-of-the-art Autoware.Universe software. 
    \item We provide FuzzSense open-source to the community to utilize in fuzzing testing/research [\href{https://anonymous.4open.science/r/FuzzSense-E680/README.md}{FuzzSense}]. 
\end{itemize}





\section{FuzzSense}

FuzzSense involves the following key components: the fuzzing broker, the fuzzing environment, and the repository. The interactions of these key components with the ADS and simulator are displayed in \figurename~\ref{fig:HL-Architecture_fuzz}.

\subsection{Fuzzing Broker}
The Fuzzing Broker is the central part of the FuzzSense framework, acting as an intermediary layer, facilitating communication between the simulator, ADS, and fuzzing environment. The fuzzing broker has full control over the exchanged sensor data and listens to data, such as steering commands.
While the Fuzzing broker was described as an intermediary for the whole framework, it additionally functions as a controller, initiating and terminating the operations in the connected Simulator and ADS.
Depending on the used Fuzzers, Simulator, and ADS, the Fuzzing Broker transforms the sensor data to the required formats of the endpoints.




\subsection{Fuzzing Environment}
 \begin{figure}[t!]
    \centering    \includegraphics[width=0.35\textwidth]{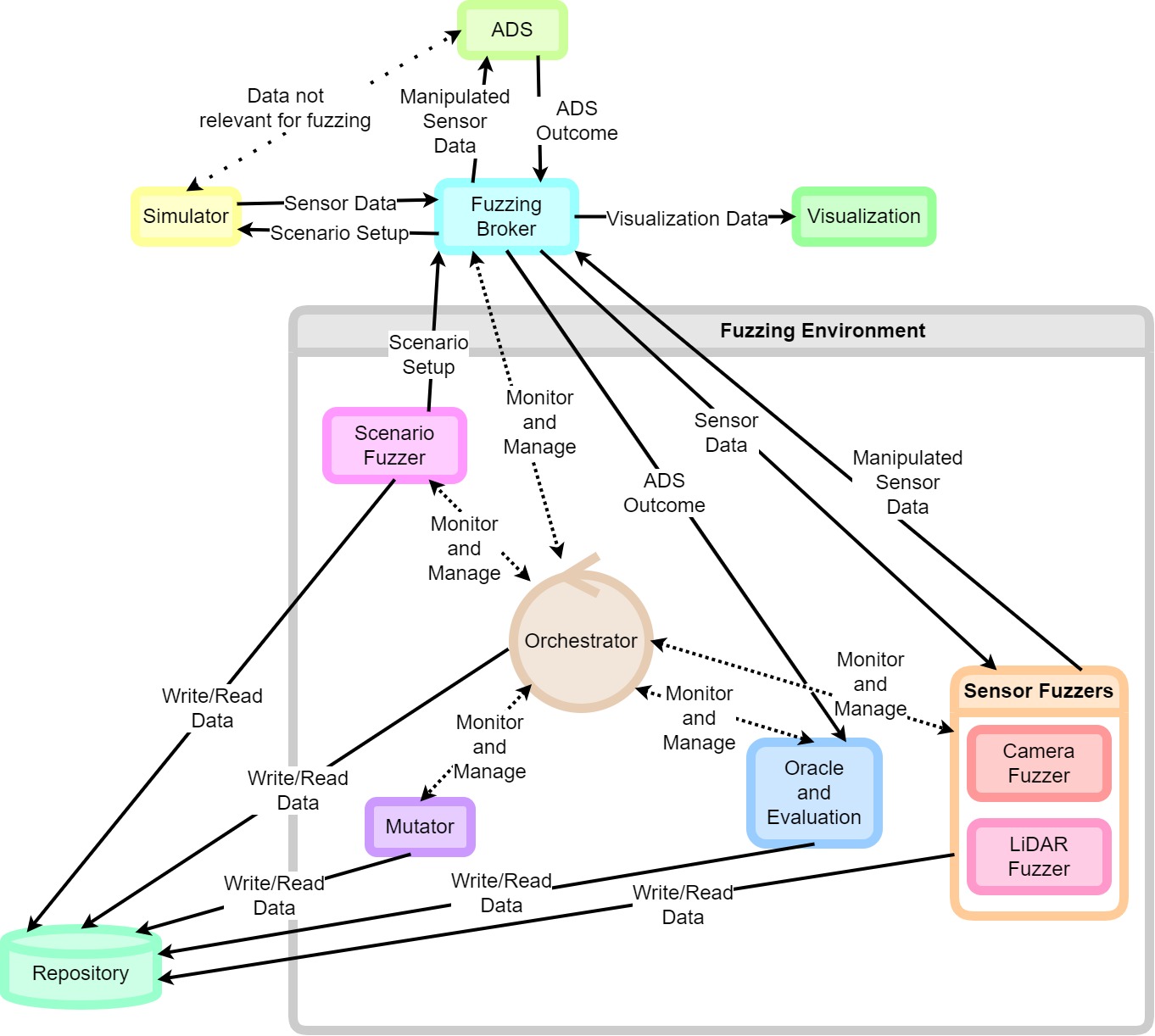}
    \caption{FuzzSense: High-level Architecture of Fuzzing Framework}
    \label{fig:HL-Architecture_fuzz}
\end{figure}
The Fuzzing Environment is the collection of the components responsible for fuzzing and creating scenarios, manipulating the sensor data, interpreting the results, and mutating parameters. This continues the idea of the modular architecture of the fuzzing framework. It also allows for the decomposition of other integrated modules, as the Mutator is not required to be a part of the fuzzers.
The Fuzzing Environment contains the following modules: orchestrator, mutator, scenario fuzzer, sensor fuzzer/s, and oracle and evaluation. 
\paragraph{Orchestrator:} The Fuzzing Environment is a composition of diverse components with unique tasks. The role of the orchestrator is to provide a central management function to ensemble these diverse components to achieve the task of fuzzing the selected targets. The idea of a fuzzing orchestrator performing a central management role was inspired by EnFuzz\cite{chen2019enfuzz}, which uses a similar design to integrate and manage diverse fuzzing modules using diverse techniques.
The Orchestrator possesses the intelligence in the Fuzzing environment.
This is reached by always knowing the current status of the fuzzing campaign and its iterations, therefore, it can start fuzzing iterations, telling each component (Fuzzers, Oracle, Mutators, Fuzzing Broker) when they should perform which of their tasks, monitor the components to understand their status to be able to efficiently start the next step with the required components.
This requires the Orchestrator to use adapters to communicate to the APIs of the different fuzzing modules. As such, no inter-communication is required for different fuzzing modules; hence, this communication is managed centrally by the Orchestrator. The benefit of central management is that expected new fuzzing modules can be integrated in less time and with less complexity. Further, it even allows decoupling the mutation of parameters and the fuzzers where they are processed.





\paragraph{Mutator:}
The Mutator creates the parameters utilized by the scenario and sensor fuzzing modules. In the first round/s the Mutator is providing the fuzzers with the seeds but does no actual mutation on them. In this architecture, the Mutator is extracted from the scenario and sensor fuzzers. The aim is to allow the combination of different mutation algorithms and fuzzers. Furthermore, it allows a closer synchronization between the mutation of parameters when using multiple fuzzers. For the proof of concept, the mutation is a brute-force/grid search iteration through parameters, where limits are applied and derived from logical boundaries like the perception distance of the sensors.



\paragraph{Scenario Fuzzer:}
Scenario fuzzers use parameters of the driving scenario as the seed pool. These can include weather, pedestrians, and other vehicles. Mutations can be built from the mission, weather, and scenario actors. Prominent scenario fuzzers include only the distinct module creating the scenarios based on parameters given by the Mutator, which is called the Scenario Fuzzer in the FuzzSense architecture. 



\paragraph{Sensor Data Fuzzers:}
Autonomous vehicles can use a range of sensor technologies and different hardware and software configurations and can be positioned at different locations on the vehicle. In general, the sensor data of any of those sensors could be fuzzed. The idea motivating our ensemble fuzzing design is that a dedicated sensor data fuzzing plug-in is responsible for each sensor data stream that should be fuzzed. The parameters provided by the Mutator can either be synchronized between several or be mutated individually.



\paragraph{Oracle and Evaluation:}
The Oracle and the Evaluation are giving further intelligence to the Fuzzing Environment. The Oracle and Evaluation component is responsible for creating ground truth, known commonly as the Golden Run. Afterward, every fuzzing iteration must be checked for possible findings, and thus, the Mutator must be provided with an evaluation of the parameters. This framework does not suggest certain conditions once a finding is detected. The idea is to set this based on the subject of testing. For instance, it could be limited to deviations of the trajectory of the Golden Run or only focus on temporal aspects (speed of the vehicle, etc.) introduced by the fuzzing.


\subsection{Repository}
In this architecture the repository enables the Fuzzing Environment to write logs, store data and dependent on the communication allow the components to exchange data. When the modules exchange data using the repository, it allows a decoupling and a simpler integration of new components, especially, because the orchestrator is handling the management centrally and thus modules do not need custom integrations with all other required in the Fuzzing Environment.

\section{Sensor Data Fuzzing}\label{sec:Approach-sensorDataFuzzing}
AD software relies on sensing data for situational awareness and to inform navigation and motion-planning activities. 
\textit{FuzzSense} is designed to apply manipulations to the sensor data stream before it reaches the downstream AD software. 
The initial version of the fuzzer manipulates pixels in the camera feed or points in the LiDAR feed. 
The fuzzer is triggered during a scenario simulation. 
For each future scenario, the fuzzing test case is mutated based on evaluation of the feedback. 
The delivery of the manipulation of the sensor data is achieved through the application of changing or adding data in the data stream based on the coordinates given by the \textit{fuzzing mask}.

\subsection{Fuzzing Mask}\label{subsec:approach-FuzzingMask}

The \textit{fuzzing mask} is created based on parameters given by the sensors and vehicle that are to be tested.
For the camera stream, which can be represented as a matrix with definitions of each pixel's coordinates, color, and sometimes the alpha channel, the \textit{fuzzing mask} provides a collection of coordinates for pixels that are changed in the camera stream.
For LiDAR, the same concept is used to add points to the point cloud, and only the distance is added. Our goal is to achieve several advantages with this approach. First, the same mutation strategy for most parameters can be used. Second, the computation of the next data points to manipulate in the LiDAR data stream is independent of the actual point cloud data. This could potentially increase the performance. Third, by limiting the space of possible manipulations in the search space, possible mutations of the parameters can be drastically reduced to the areas of interest (e.g., in front of the vehicle). Within a point cloud, points can be hidden behind others from the sensors perspective. The concept with the \textit{fuzzing mask} prevents such cases so that no added points are shadowed by other added points (see \figurename~\ref{fig:fuzzingMaskLiDAR}).

\begin{figure}[t!]
    \centering
    \includegraphics[width=0.99\columnwidth]{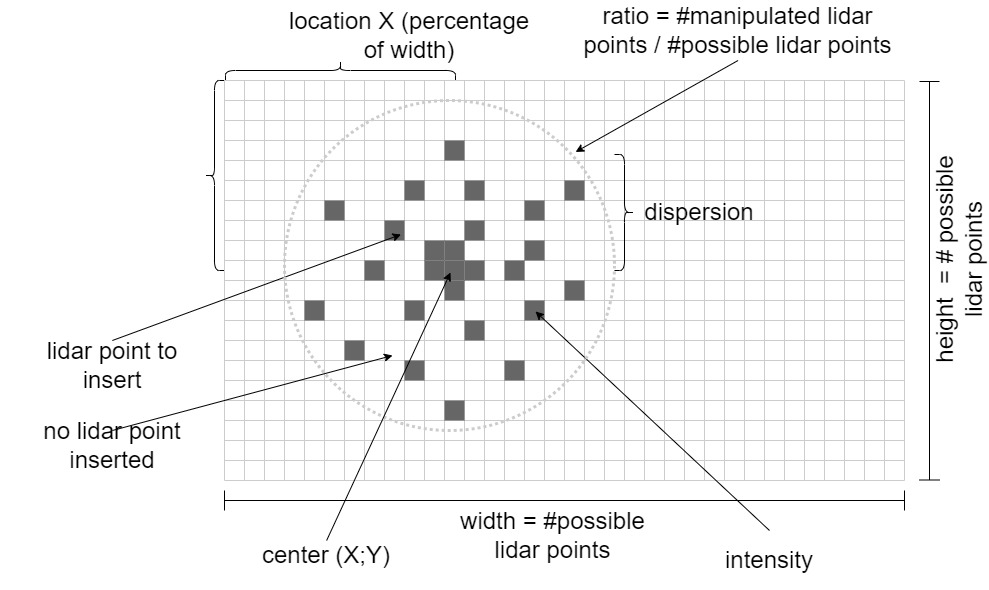}
    \caption{Fuzzing Mask for LiDAR.}
    \label{fig:fuzzingMaskLiDAR}
\end{figure}

 The fuzzing mask $\mathcal{F}$(Algorithm~\ref{alg:GenerateFuzzingMask}) is defined as a set of coordinates where the sensor data is manipulated $\mathcal{F} = \{ (x_i, y_i) \mid x_i \in [0, W], y_i \in [0, H]\}$. For the camera sensor, the location of the pixel, and for the LiDAR sensor corresponds to the coordinates of a perpendicular plane in the pointcloud where each point is inserted. The third dimension for LiDAR is provided by the distance parameter between the LiDAR sensor and the plane. The coordinates are relative to width, height, and, for 3D data, the center of the plane. For the camera stream, they are taken from the metadata of the sensor stream, and for LiDAR, they are preset and could potentially be mutated.


\input{Algorithms/Algorithm_1}

 Let $r_f$ represent the fuzzing change ratio, defined as $r_f = \frac{N_c}{W \times H}$. 
 Where: $N_c$ is the number of changed data points, $W$ and $H$ are the width and height of the fuzzing mask matrix in discrete steps (e.g. pixels for the camera stream). 
 The result is expressed as a percentage.
 Then, let $\sigma_f$ represent the standard deviation of the manipulated data points, computed as the deviation relative to width $W$ and height $H$. Together, $X$ and $Y$ are the coordinates of the center of the fuzzing mask and the means of the standard-deviation. $\overline{x}$ and $\overline{y}$ are the vectors corresponding to the each $x$ and $y$ coordinate vector respectively. In line 3-4 $W$ and $H$ ensure, that no coordinates outside of the fuzzing mask are created. 
 Where in line 6 $\mathcal{F}$ is created by column stacking the $\overline{x}$ and $\overline{y}$ arrays with the calculated normal-distributions.

\subsection{Multi-Stage Approach}\label{sec:multiplestages}
FuzzSense combines multiple stages during fuzzing. Each time the fuzzing setup is started, it is called a \textbf{Fuzzing Campaign}. Each of the scenarios running with different fuzzing parameters is defined as a \textbf{Fuzzing Iteration}. This allows to better distinct between phases and to have an easier understanding of the complete process and architecture. The aim of this process design choice is that the focus for the fuzzing campaign can be chosen with more granularity as the multi-stages allows to provide intelligence to the iterations. The logic when to exit the inner iteration (sensor fuzzing iteration) can be set based on the aim of the fuzzing campaign. This is possible, because the inner and outer iteration (scenario fuzzing iteration) can be logically separated.

\subsubsection{Fuzzing Campaign}
The Fuzzing Campaign defines the whole duration of the fuzzer running. A Fuzzing Campaign consists of one or many Fuzzing Iterations. To start a fuzzing campaign, one or several seeds are required. Each seed contains starting values for each parameter. While there is not any condition met, which qualifies the end of the campaign, new scenario fuzzing iterations are started. The campaign also could be stopped manually. The final step is to stop all required services and store the results from the fuzzing campaign to allow further investigations.


\subsubsection{Fuzzing Iteration}\label{subsec:FuzzingIteration}
The Fuzzing Iteration defines one single scenario run. It starts with the parameter mutation and ends once the scenario is stopped because of a failure or because it has successfully finished. The fuzzing of every single data frame is not called iteration. A here defined Fuzzing Iteration includes all those manipulated sensor data frames throughout the whole scenario until it finishes or fails with a finding. As the main focus of the fuzzing is on the sensor data, the mutation for the scenario parameters is not performed in every iteration. Thus, the same scenario is present throughout several iterations. To distinguish also between those two, there can be \textit{Scenario Fuzzing Iterations} and \textit{Sensor Fuzzing Iterations}. One Scenario Fuzzing Iteration consists of one or many Sensor Fuzzing Iterations.

\paragraph{Scenario Fuzzing Iteration}
The ADS of the AV must act within a scenario to allow relations to its intended real-world use. A scenario defines not only the ego-vehicle itself but also the road, traffic signs, and signals, road conditions, environment, other actors, including their behavior, and the weather conditions.
The Scenario Fuzzing Iteration is the outer iteration and contains all Sensor Fuzzing Iterations in the same scenario. It contains the following steps: 
\begin{description}
  \setlength{\itemsep}{-0.1em}
  \item[Step 1:] Mutate Scenario Parameters
  \item[Step 2:] Create a Scenario and set it up in the simulator and ADS
  \item[Step 3:] Create Golden Run
  \item[Step 4:] Start \textit{Sensor Fuzzing Iterations}
\end{description}


\paragraph{Sensor Fuzzing Iteration}\label{subsubsec:sensorFuzzingIteration}
Within the same Scenario Fuzzing Iteration, the parameters for the Fuzzing Mask should not be the same twice. However, within a new Scenario Fuzzing Iteration, the same parameters can be used again. Each sensor fuzzer takes the original sensor data from the simulator and applies manipulations to the data stream before it reaches the ADS. Those manipulations are single pixels in the camera feed or points in the LiDAR feed. In the current state, within one run, the planned drive of the vehicle, no mutations on the parameters are performed. This means the same fuzzing masks are applied to the data streams from the start to the end of the drive. The mutator is only active between runs. Therefore, compared to a plain simulation, the only computational overhead during a running simulation is the rerouting and manipulation of the sensor data. It contains the following steps: 
\begin{description}
  \setlength{\itemsep}{-0.1em}
  \item[Step 1:] Mutate Sensor Parameters
  \item[Step 2:] Set scenario up in simulator and ADS
  \item[Step 3:] Create Fuzzing Masks 
  \item[Step 4:] Start scenario and manipulate sensor data streams
\end{description}




\section{Experiment \& Results}
\subsection{Experimental Setup}
The evaluation of FuzzSense is conducted in AWSIM, a high-fidelity, digital-twin simulation environment. The target AD system uses the Autoware.Universe software framework. As this instantiation of the AD software uses the LiDAR sensor for perception and localisation, the sensor fuzzing module is configured to fuzz the LiDAR sensor.
The evaluation was conducted on a system running Ubuntu 22.04.03 LTS with 1 TB of storage, 32 GB of CPU memory, 10 GB of GPU memory, a 12th Gen Intel® Core™ i7-12700KF processor, and a GeForce RTX 3080 Lite Hash Rate graphics card.

\subsection{Results \& Discussion}\label{sec:findings}

The driving scenario consists of a planned navigation in an urban driving environment. We selected an urban environment since attacks can cause more severe effects within a congested operational driving domain. As the vehicle navigates through its planned trajectory, the sensor fuzzing plug-in of FuzzSense initiates its fuzzing mask, manipulating the parameters of the LiDAR 3D geometry. For this set of experiments, the parameters were randomly set at x (0.4),y (0.5), the distance of the fuzzed LiDAR points (30m), and the intensity 0.1. and dispersion (width 100, height 60). The experiments mutated the location and dispersion parameters. The fuzzing broker is fuzzing every frame. In the simulation, the environment exhibits a performance of time of approx. 30 frames per second or 33 milliseconds. \figurename~\ref{fig:finding_3-wallDistance1OutOfWay} displays the initiation of the fuzzing mask (the yellow box is used for identification and does not represent the full mask) to the driving simulation. The fuzzing mask is applied at different distances from the vehicle and different locations within the environment. As shown in \figurename~\ref{fig:finding_3-wallDistance1OutOfWay}, the fuzzing mask is located at an approaching distance to the vehicle of approx. 30 meters outside the lane does not produce any unsafe changes in the vehicle the vehicle's behavior. 


\begin{figure}[h!]
    \centering
    \includegraphics[width=0.45\textwidth]{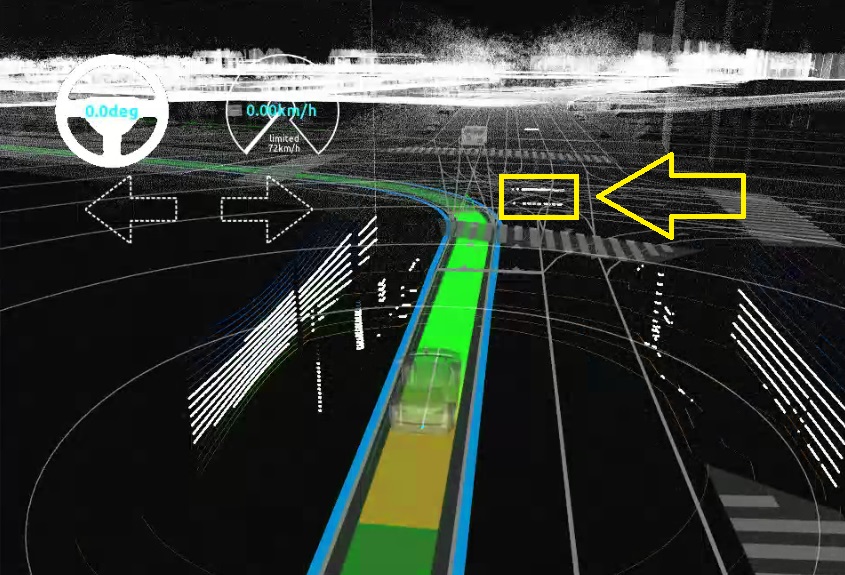}
    \caption{Fuzzing Mask applied to the right edge of lane}
    \label{fig:finding_3-wallDistance1OutOfWay}
\end{figure}


\figurename~\ref{fig:finding_4-wallDistance1InWay_marked} displays the movement of the fuzzing mask to a more central location in the driving environment. The fuzzing parameters for amount and dispersion are the same as \figurename~\ref{fig:finding_3-wallDistance1OutOfWay} in both fuzzing iterations. The parameter for the distance is the same for both. 
\begin{figure}[h!]
    \centering
    \includegraphics[width=0.45\textwidth]{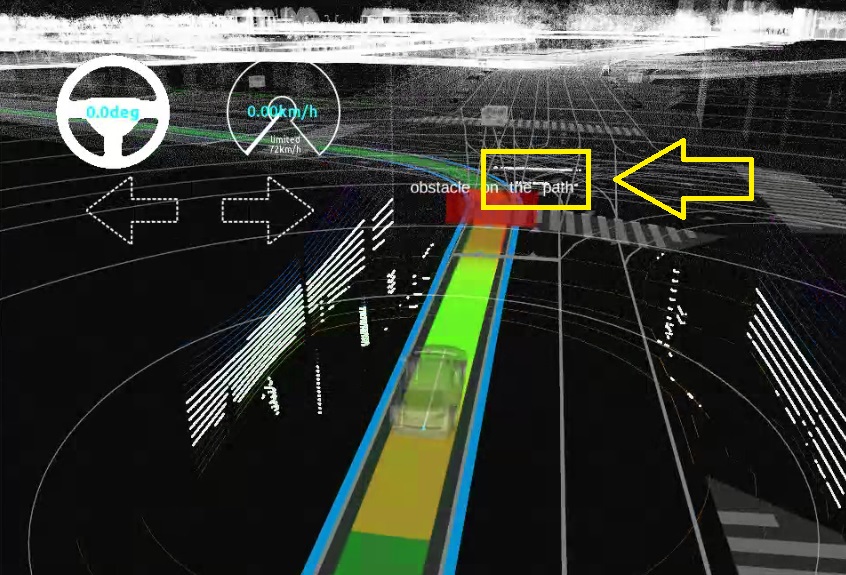}
    \caption{Fuzzing Mask applied to central location of vehicle trajectory}
    \label{fig:finding_4-wallDistance1InWay_marked}
\end{figure}
The affect of the fuzzing mask displayed in \figurename~\ref{fig:finding_4-wallDistance1InWay_marked}, is that the vehicle detects the fuzzed LiDAR points as an obstacle (red wall) and plans a reduction in acceleration to observe the obstacle. This can be seen by the orange color in the planned trajectory.



\figurename~\ref{fig:finding_5-wallDistance2OutOfWay} displays the fuzzing mask applied at a close distance and within the planned trajectory of the vehicle. The vehicle detects the fuzzing mask as an object in immediate proximity to the vehicle and therefore initiates a braking action. The vehicle is unable to recompute an alternative planned trajectory due to the fuzzed points presenting an obstacle across the road and therefore the vehicle is unable to progress. 

\begin{figure}[h!]
    \centering
    \includegraphics[width=0.35\textwidth]{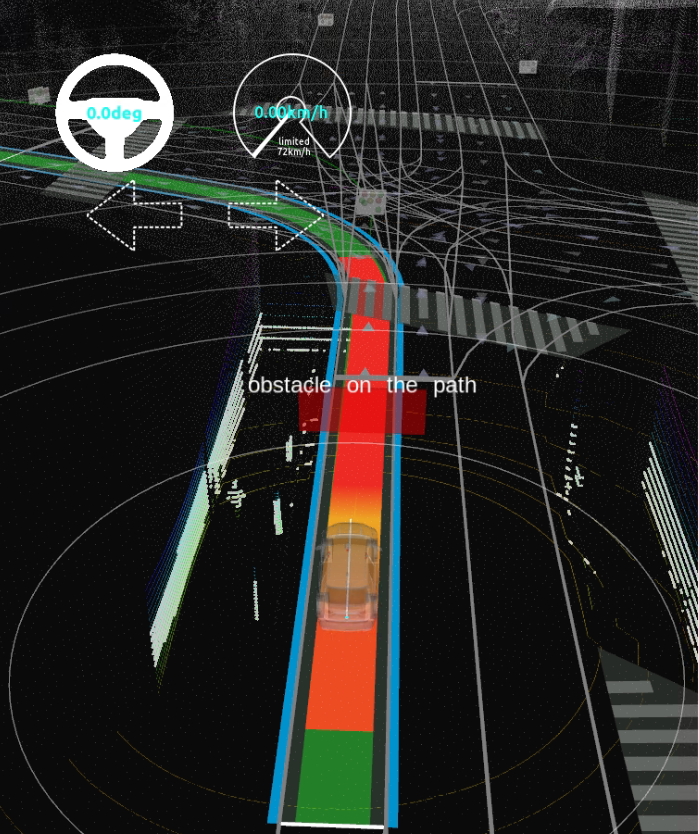}
    \caption{Top down view of vehicle with fuzzing mask affecting planned navigation of the vehicle}
    \label{fig:finding_5-wallDistance2OutOfWay}
\end{figure}



The experiments provide initial feedback on the utility of FuzzSense. From observing the behaviour of the AD software, displayed in Figures~\ref{fig:finding_4-wallDistance1InWay_marked} and \ref{fig:finding_5-wallDistance2OutOfWay} we can discern that sensor fuzzing is a useful exercise to find vulnerabilities of the AD software stack. The results indicate that the AD software is either unstable or can be influenced by inserted LiDAR points. We found that when the fuzzing mask was located on or near the planned trajectory of the vehicle, the perception algorithm was unable to filter the manipulated points and instead, observed them as an obstacle. Further to this, when the fuzzing mask was located in close proximity to the vehicle, it resulted in a complete stop of the vehicle.

\section{Related Work}
The EnFuzz architecture~\cite{chen2019enfuzz} demonstrates the advantage of combining multiple fuzzers which use diverse techniques of fuzzing, to get a greater and deeper penetration of the target. The EnFuzz design further inspired our adoption within FuzzSense of an orchestrator (monitor) for coordination. Our contribution is unique from EnFuzz as our focus is specific to AD software and we incorporate in the design considerations for the diversity of AD technology and targets.

Aforementioned, there are various fuzzers focused on disparate targets of the AD system. Popularly cited fuzzing tools include DeepRoad~\cite{10.1145/3238147.3238187}, DeepTest 
~\cite{10.1145/3180155.3180220} which target the camera sensor and AV-Fuzzer~\cite{9251068}, Auto-Fuzz~\cite{10.1109/TSE.2022.3195640} and DriveFuzz~\cite{10.1145/3548606.3560558} which target the driving scenario. These fuzzers are not designed to operate concurrently with different fuzzers, but focus on a seed pool limited to there target. For the optimization of the search space reduction, these fuzzing tools mainly focus on driving quality and task performance metrics as a measure to direct the mutations towards more promising scenarios where the ego-vehicle is more likely to struggle. 

Our work does not aim to compete with these fuzzers nor do we seek to build on there designers. FuzzSense, is an overarching framework whose concept is based on enabling the usage of the fuzzing tools as plug-ins in an integrated fuzzing environment. A future test case would be to use DeepRoad~\cite{10.1145/3238147.3238187} and DriveFuzz~\cite{10.1145/3548606.3560558} within FuzzSense to understand how diverse fuzzing techniques generate bugs.     



 
\section{Conclusion \& Future Work}

In this work we presented our idea for the design of an ensemble fuzzing framework for AD software, which we call FuzzSense. FuzzSense is designed for vehicles with several different sensors. The architecture consists of many different modules which perform tasks fulfilling each a crucial part in the execution of fuzzing and of the coordinating and monitoring of those diverse fuzzing modules. We developed a plug-in for fuzzing LiDAR point clouds and utilised to find vulnerabilities in Autoware.Universe AD software. Future work, aims to experiment with FuzzSense utlising the modularity to benchmark the performance of different fuzzing plug-ins. Further, advancing the design of the fuzzing mask by adding support for further sensor types. As part of providing FuzzSense open-source, we also aim to actively gather community feedback and develop the framework further.

\addcontentsline{toc}{chapter}{Bibliography}
\bibliographystyle{ieeetr}      
\bibliography{references}

\end{document}

%% file: Algorithms/Algorithm_1.tex
 \begin{algorithm}
 \caption{Generate Fuzzing Mask $\mathcal{F}$}
 \label{alg:GenerateFuzzingMask}
 \begin{algorithmic}[1]
 \Require $r_f, \sigma_f, X, Y, W, H$
 \State $(\sigma_x, \sigma_y) \gets (W*\sigma_f,  H*\sigma_f)$
 \State $r_f \gets W*H*r_f$
 \State $\overline{x} = \mathcal{N} (r_f, \sigma_x, X, W )$
 \State $\overline{y} = \mathcal{N} (r_f, \sigma_y, Y, H ) $
 \For{$i \gets 0$ to $r_f-1$}
     \State $\mathcal{F} \gets add (x[i],y[i])$
 \EndFor
 \State \Return $\mathcal{F}$
 \end{algorithmic}
 \end{algorithm}
